# Transportation dynamics of dielectric particles with the gradient forces in the field of orthogonal standing laser waves

A.A. Afanas'ev[1], V.M. Volkov[2], Yu.A. Kurochkin[1], D.V. Novitsky[1*]

[1]B.I. Stepanov Institute of Physics, National Academy of Sciences of Belarus, Nezavisimosti Avenue 68, 220072 Minsk, Belarus

[2]Mechanics and Mathematics Faculty, Belarusian State University, Nezavisimosti Avenue 4, 220030 Minsk, Belarus

*Corresponding author: dvnovitsky@gmail.com

We develop the theory of transportation and localization of a transparent dielectric spherical particle with the gradient forces in the interference field of orthogonally directed standing laser waves $E_z(\cos kz)$ and $E_x(\cos kx)$. It is shown that, when the waves $E_z$ and $E_x$ are coherent, the interference radiation field contains two harmonic components with the periods $\Lambda_0 = \pi/k$ and $\Lambda_\Delta = \pi/(k\sin(\pi/4))$. The amplitudes of the gradient force components depend on the ratio of the particle radius $R$ to the modulation periods due to inhomogeneity of radiation in the particle volume and are given by the Bessel functions $J_{3/2}(2\pi R/\Lambda_0)$ and $J_{3/2}(2\pi R/\Lambda_\Delta)$. We find the critical particle radii $R_0$ and $R_\Delta = \sqrt{2}R_0$ defined by the Bessel functions zeros and corresponding to the vanishing components of the gradient forces. In particular, for the radiation with the wavelength $\lambda_0 = 1.064$ μm and a particle in water, the smallest critical radii are $R_0 = 0.286$ μm and 0.492 μm and $R_\Delta = 0.404$ μm and 0.696 μm, respectively. For a number of special cases, we obtain the analytical solutions of the Newton equations and the particle trajectories that depend on the ratio of wave intensities and the particle radius. The results can be used to study the dynamics of the "optical assembly" of a two-dimensional particles matrix which behaves as a molecular crystal [Mellor and Bain, Chem. Phys. Chem. 7 (2006) 329-332].

## 1. Introduction

It is known [1-3] that in spatially inhomogeneous fields of laser radiation, there is the gradient component of the light pressure force $\vec{F}_\nabla$ acting on polarizing particles along the intensity gradient. Depending on the optical properties and particle size, wavelength, and scale of the spatial inhomogeneity of laser radiation, small particles are pulled into the intensity maxima or pushed out of them under the action of the force $\vec{F}_\nabla$ [4-8]. From a practical point of view, the effect of gradient forces on liquid suspensions of dielectric particles in periodically modulated radiation created by the interference of two or more laser beams is of particular interest [9-15]. Under the action of gradient forces, the concentration dynamic lattices are induced in the suspension, the establishment and decay times of which are determined by the properties of particles and radiation, as well as by the viscosity of the surrounding liquid. Despite the fact that each component of the suspension (the particles and surrounding liquid) does not exhibit nonlinear optical properties, such an artificially created heterogeneous medium is a highly efficient broadband nonlinear material for continuous laser radiation of relatively low power [7]. For example, in the study of four-wave mixing of argon laser radiation ($\lambda_0 = 5145$ Å) in the water suspension of latex microspheres with radius $R = 1.17$ μm and concentration $N_0 = 6.5 \cdot 10^{10}$ cm$^{-3}$, the optical Kerr coefficient $n_2$ was measured to be $3.6 \cdot 10^{-9}$ cm$^2$/W [16], that is $10^5$ times larger than that of carbon disulfide (CS$_2$). In [7, 17] the theory of four-wave mixing was developed in the diffusion limit when a liquid suspension of dielectric particles is similar to a medium with phase cubic nonlinearity. Theoretical and experimental studies of stimulated concentration (diffuse) scattering were performed in [18–20]. In [21] an experiment was carried out on the formation of a two-dimensional matrix of polystyrene particles with $R =$ 150–300 nm in water under the action of gradient forces in the interference field of the radiation of two continuous lasers. In this experiment, when the radiation was reflected and refracted at the interface of the glass substrate with a drop of suspension, the particles were actually exposed to the two multidirectional gradient forces resulting in an "optical assembly" of a two-dimensional particle matrix, which had the properties of ordinary molecular crystals [21]. This result opens up the prospects for optical synthesis of three-dimensional (bulk) matrices from small particles – the "optical assembly" of crystals, which are of undoubted interest for scientific and practical applications.

In this paper, we develop the theory of transportation and localization of dielectric particles in a liquid under the action of gradient forces in the field of a pair of orthogonal standing waves of laser radiation. This theoretical model does not strictly correspond to the experimental conditions of [21]. However, for the theoretical analysis, such a problem statement is the simplest and allows one to elucidate the basic laws of the "optical assembly" dynamics for a two-dimensional matrix of small-sized particles.

## 2. Main equations

The field amplitude in the case of two orthogonal pairs of counter-propagating waves can be presented as

$$E_0 = \frac{1}{2}\left[\left(E_{+z}e^{ikz} + E_{-z}e^{-ikz}\right) + \left(E_{+x}e^{ikx} + E_{-x}e^{-ikx}\right)\right]e^{-i\omega t} + \text{c. c.}, \quad (1)$$

where $E_{\pm z}$ and $E_{\pm x}$ are the amplitudes of waves propagating along the axes $z$ and $x$, respectively, $\omega$ and $k$ are the frequency and wavenumber. Note that all the waves have the same polarization and can interfere with each other.

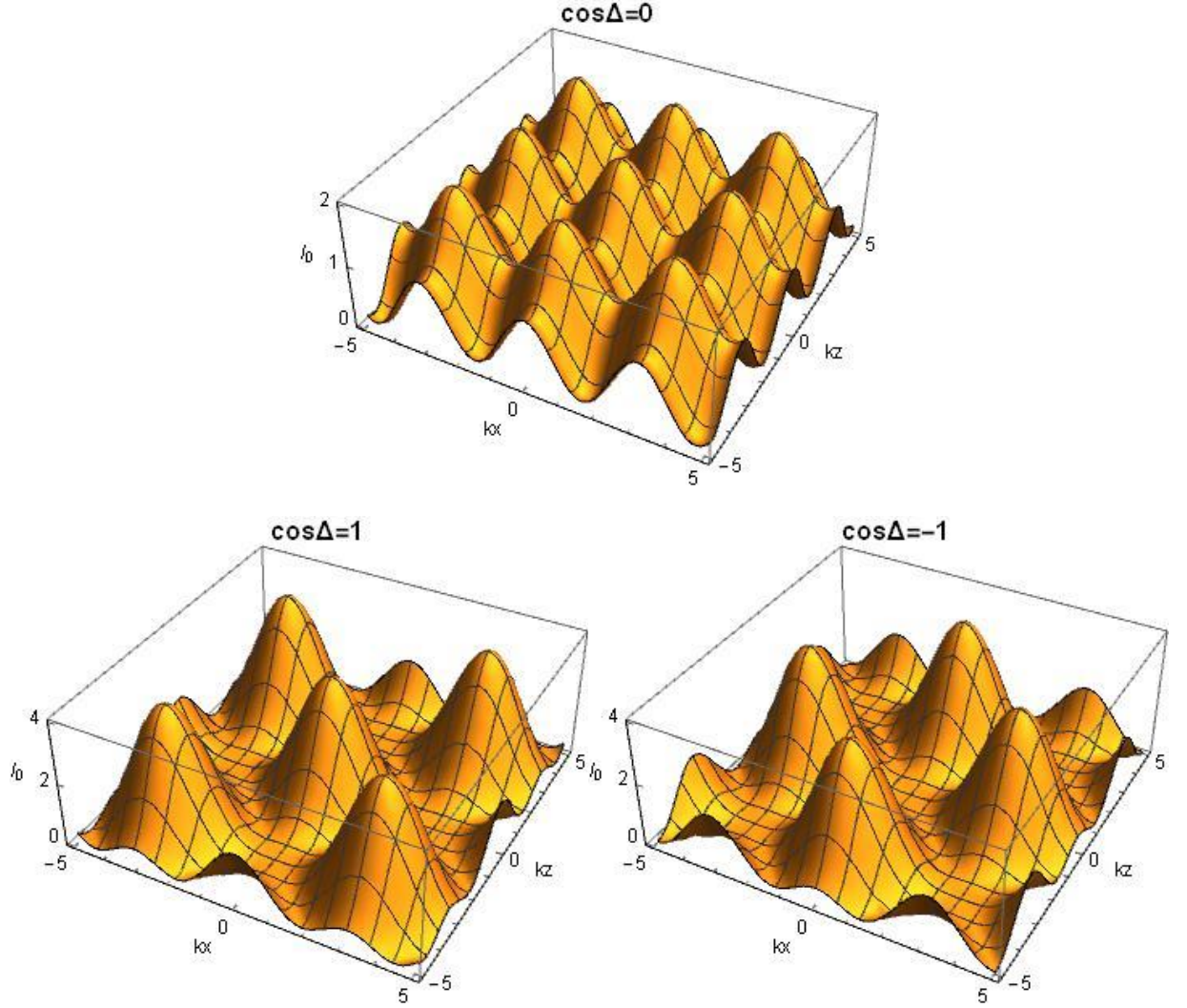

Fig. 1. Distribution of the radiation intensity $I_0$ (in arb. units) for various values of the function $\cos\Delta$.

For the equal amplitudes of counter-propagating waves $E_{\pm r} = |E_r|\exp(i\varphi_r)$ ($r = z, x$), we can derive the radiation intensity $I_r = cn/(8\pi)|E_r|^2$ from Eq. (1) as

$$I_0 = 2(I_z\cos^2 kz + I_x\cos^2 kx + 2\sqrt{I_z I_x}\cos\Delta \cdot \cos kz \cdot \cos kx), \qquad (2)$$

where $\Delta = \varphi_z - \varphi_x$ is the phase difference, $n$ is the refractive index of a liquid, in which a particle resides. The last term in Eq. (2) describes the interference of orthogonally propagating waves $E_z$ and $E_x$. Note that the case $\cos\Delta = 0$ means the absence of interference, i.e., it is the analogue for the mixing of the waves $E_z$ and $E_x$ with the random phase modulation, so that $\langle\cos\Delta(t)\rangle_t = 0$. For $\cos\Delta = 0, \pm 1$, Eq. (2) gives the simple relations:

$$I_0(z,x) = 2I_z \begin{cases} \cos^2 kz + \gamma^2 \cos^2 kx, & \text{for } \cos\Delta = 0, \\ (\cos kz \pm \gamma \cos kx)^2, & \text{for } \cos\Delta = \pm 1, \end{cases} \quad (3)$$

where $\gamma = \sqrt{I_x/I_z}$. Figure 1 shows the graph of the function $I_0(z, x)$ for $\gamma = 1$. It is seen that for $\cos\Delta = \pm 1$, the modulation periods and radiation intensity in the maxima of interference pattern are twice as much as in the case of $\cos\Delta = 0$. At the same time, when changing the sign of $\cos\Delta$, the pattern shifts by $\pi$ in both coordinates.

Taking Eq. (2) into account, the expression for the gradient force $\vec{F}_\nabla$ can be generally written as follows [4-7]:

$$\vec{F}_\nabla = \vec{F}_z + \vec{F}_x = 2\pi \frac{n}{c} \alpha \frac{1}{V} \int_V \left( \vec{z} \frac{\partial I_0}{\partial z} + \vec{x} \frac{\partial I_0}{\partial x} \right) dV, \quad (4)$$

where $\alpha$ and $V = (4\pi/3)R^3$ are the polarizability and the volume of the spherical particle with the radius $R$, $\vec{z}$ and $\vec{x}$ are the unit vectors along the cartesian coordinates ($\vec{z} \perp \vec{x}$). The integral in Eq. (4) takes into account radiation inhomogeneity in the particle volume. For small particles ($kR \ll 1$), radiation inside the particle can be considered homogeneous, so that

$$\frac{1}{V}\int_V \left( \vec{z} \frac{\partial I_0}{\partial z} + \vec{x} \frac{\partial I_0}{\partial x} \right) dV \approx \vec{z} \frac{\partial I_0}{\partial z} + \vec{x} \frac{\partial I_0}{\partial x}.$$

In this case, the polarizability $\alpha$ is determined by the expression [22, 23]

$$\alpha = \frac{\bar{m}^2 - 1}{\bar{m}^2 + 2} R^3 = \alpha_0 R^3, \quad (5)$$

where $\bar{m} = n_0/n$, $n_0$ is the refractive index of the particle material. From Eqs. (2) and (4), for a particle of arbitrary radius $R$, we have

$$F_z = -4\pi k \frac{n}{c} \alpha \{u_0(kR) I_z \sin 2kz + 2u_\Delta(kR)\sqrt{I_z I_x} \cos\Delta \cdot \sin kz \cdot \cos kx\}, \quad (6a)$$

$$F_x = -4\pi k \frac{n}{c} \alpha \{u_0(kR) I_x \sin 2kx + 2u_\Delta(kR)\sqrt{I_z I_x} \cos\Delta \cdot \sin kx \cdot \cos kz\}, \quad (6b)$$

where the functions $u_0(kR)$ and $u_\Delta(kR)$ take into account radiation inhomogeneity in the particle volume and are defined by the integrals as follows:

$$u_0(kR) = \frac{1}{V}\int_V \sin 2kr \, dV \text{ and } u_\Delta(kR) = \frac{1}{V}\int_V \sin kr \cdot \cos kr' \, dV, r \neq r'. \quad (7)$$

Keeping in mind Eqs. (2)–(6), the Newton equations for the dimensionless particle coordinates $\hat{z} = kz$ and $\hat{x} = kx$ can be written as

$$m\frac{d^2\hat{z}}{dt^2} + 6\pi R\eta \frac{d\hat{z}}{dt} = -g_0 \sin\hat{z}(u_0 I_z \cos\hat{z} + u_\Delta\sqrt{I_z I_x}\cos\Delta \cdot \cos\hat{x}) \equiv F_z(\hat{z},\hat{x}), \quad (8a)$$

$$m\frac{d^2\hat{x}}{dt^2} + 6\pi R\eta \frac{d\hat{x}}{dt} = -g_0\sin\hat{x}(u_0 I_x \cos\hat{x} + u_\Delta\sqrt{I_z I_x}\cos\Delta \cdot \cos\hat{z}) \equiv F_x(\hat{z},\hat{x}), \quad (8b)$$

where $m = V\rho$ is the mass of the particle made from the material with the density $\rho$, $6\pi R\eta$ is the friction coefficient in the liquid with the dynamic viscosity $\eta$, $g_0 = 8\pi k^2 n\alpha/c$. Obviously, for $\cos\Delta = 0$, each component of the gradient force depends only on a single («its own») coordinate ($F_z(z)$ and $F_x(x)$) and Eqs. (8) become two independent equations.

### 3. Analysis of the functions $u_0(kR)$ and $u_\Delta(kR)$

A characteristic parameter of integrals (7) is the ratio of the particle radius $R$ to the period of radiation modulation $\Lambda$ ($2\pi R/\Lambda$) [6]. The first terms in the right-hand sides of Eqs. (6) appear due to interference of the counter-propagating waves $E_{+z}E_{-z}$ and $E_{+x}E_{-x}$ and describe the radiation modulation with the period $\Lambda_0 = \pi/k$ along the axes $z$ and $x$, respectively. The second terms are the result of interference of the orthogonal waves $E_{+z}E_{\pm x}$ and $E_{-z}E_{\pm x}$ and modulate radiation with the period $\Lambda_\Delta = \pi/\left(k\sin\left(\frac{\pi}{4}\right)\right)$ in the directions rotated by the angle $\pi/4$ with respect to the axes $z$ and $x$. Having in mind these remarks and the relation $\Lambda_\Delta/\Lambda_0 = \sqrt{2}$, we calculate the integrals (7) and obtain the results as follows:

$$u_0(kR) = 3\sqrt{\frac{\pi}{2}}(2kR)^{-3/2}J_{3/2}(2kR); \quad u_\Delta(kR) = 3\sqrt{\frac{\pi}{2}}(\sqrt{2}kR)^{-3/2}J_{3/2}(\sqrt{2}kR), \quad (9)$$

where $J_{3/2}(kR)$ is the Bessel function; $2kR = 2\pi R/\Lambda_0$, $\sqrt{2}kR = 2\pi R/\Lambda_\Delta$. One can show that for small particles (with $kR < 1$), when radiation inhomogeneity in the particle volume can be neglected, we have $u_0 \approx u_\Delta \approx 1$. Graphs of the functions $u_0(kR)$ and $u_\Delta(kR)$ are given in Fig. 2. At the points of two smallest zeros of the Bessel function, $J_{3/2}(2kR_0) = 0$ ($kR_0 = 2.247$ and $3.863$), we have $u_0(kR_0) = 0$. At these points, for $\cos\Delta = 0$, the gradient force is $\vec{F}_\nabla = 0$ ("zero force effect") [7] regardless of the particle position with respect to the field interference pattern. Accordingly, in this case, for certain values of $R_0/\Lambda$, the particle can be in a steady state at the maximum or minimum of intensity and does not feel the force at all, remaining at rest.

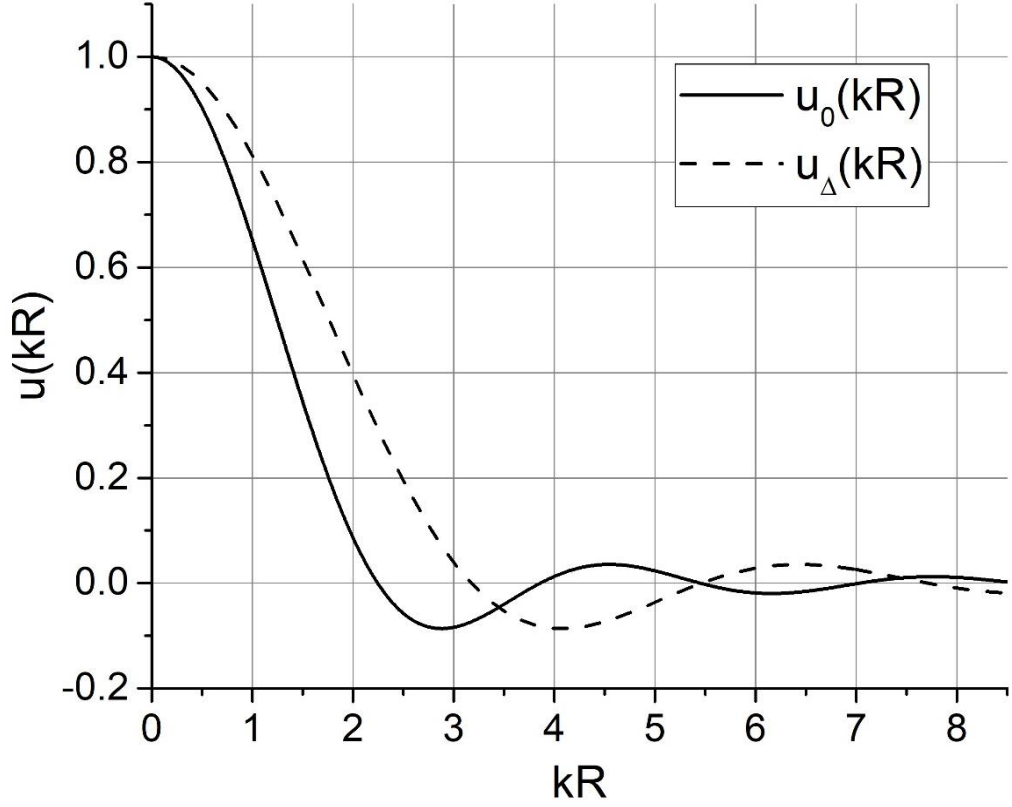

Fig. 2. Graphs of the functions $u_0(kR)$ and $u_\Delta(kR)$.

In Lekner's quantum-mechanical theory of light scattering on a spherical particle [8] and the corresponding calculations of the gradient force in the standing wave, "the zero-force effect" is reached for the smallest radii determined by the values $kR_0 = 2.445$ and $3.985$. One can see that our values of $kR_0$ are 8% and 3% less than those obtained in [8]. In order to obtain $R_0$ from Lekner's theory [8], one has to calculate the sums of infinite series. On the contrary, we have obtained the very simple relations with $R_0$ given by the zeros of the tabulated Bessel function $J_{3/2}(2kR)$ [24], so that there is no need for any additional calculations. Estimates for the dielectric particles in water ($n = 1.33$) in the radiation field with the wavelength $\lambda_0 = 1.064$ μm ($\Lambda_0 = \lambda_0/(2n) = 0.4$ μm) result in the values as follows: $R_0 = 0.286$ μm and $0.492$ μm. Moreover, the roots of the equation $u_\Delta(kR_\nabla) = 0$, as noted above, are numerically larger than those of the equation $u_0(kR_0) = 0$, namely $R_\nabla = \sqrt{2}R_0$ and, hence, $R_\nabla = 0.404$ μm and $0.696$ μm.

### 4. Analysis of the truncated Newton equations

Due to the small mass of the particle, the second derivatives in Eqs. (8) can be neglected in a wide range of radiation intensities and the truncated equations can be used [25]:

$$\frac{d\hat{z}}{dt} = -gI_z \sin\hat{z}(u_0\cos\hat{z} + \gamma u_\Delta\cos\Delta \cdot \cos\hat{x}), \quad (10a)$$

$$\frac{d\hat{x}}{dt} = -gI_x\sin\hat{x}(u_0\cos\hat{x} + \frac{1}{\gamma}u_\Delta\cos\Delta \cdot \cos\hat{z}), \tag{10b}$$

with the initial conditions $\hat{z}(t=0) = \hat{z}_0$ and $\hat{x}(t=0) = \hat{x}_0$; $g = 4k^2 n\alpha/(3cR\eta)$.

It follows from Eqs. (10) that for both components of the gradient force, we have $F_z = F_x = 0$, if at least one of the following conditions for the particle coordinates is satisfied:

$$\sin\hat{z} = \sin\hat{x} = 0, \tag{11a}$$

$$\cos\hat{z} = -\gamma\cos\hat{x} \quad \text{at } u_\Delta\cos\Delta/u_0 = 1. \tag{11b}$$

Thus, if the initial coordinates $\hat{z}_0$ и $\hat{x}_0$ satisfy one of the relations (11), then the particle will remain at rest.

In general, the solution of Eqs. (10) can be found only with numerical methods. Therefore, below we consider some special cases that allow analytical solutions.

### 4.1. The case of $\cos\Delta = 0$

As noted above, for $\cos\Delta = 0$, Eqs. (10) become two independent equations with the solutions of the form,

$$\tan\hat{z}(t) = \tan\hat{z}_0 \cdot e^{-t/\tau_z} \text{ and } \tan\hat{x}(t) = \tan\hat{x}_0 \cdot e^{-t/\tau_x}, \tag{12}$$

where $\tau_z = 1/(G_0 I_z)$ and $\tau_x = 1/(G_0 I_x) = \tau_z/\gamma^2$ are the characteristic times of particle transportation to the closest intensity maximum (minimum) depending on the initial coordinates; $G_0 = gu_0$. In Fig. 3, the particle trajectories are shown for $\hat{z}_0 = \hat{x}_0 = 0.9\pi/2$ at different values of the parameter $\gamma^2$ as calculated from Eqs. (12) for $G_0 > 0$. It is seen that for $\gamma^2 = 1$, the trajectory is the rectangle diagonal on the plane $(\hat{z}, \hat{x})$. For $\gamma^2 \neq 1$, the trajectory is not a straight line due to the difference in the particle's velocity components, $v_z(t) \neq v_x(t)$. For example, for $\gamma^2 = 2$ ($I_x > I_z$), in the initial moments of time $\tau_z \leq 2$ one has $v_x > v_z$, whereas later ($\tau_z \geq 2$) the relation between the velocity components becomes inverse ($v_x < v_z$). For $\gamma^2 = 0.5$, the situation is the opposite. Thus, for $\gamma^2 \neq 1$, a change in the $v_z(t)/v_x(t)$ ratio leads to a curvature of the particle's trajectory.

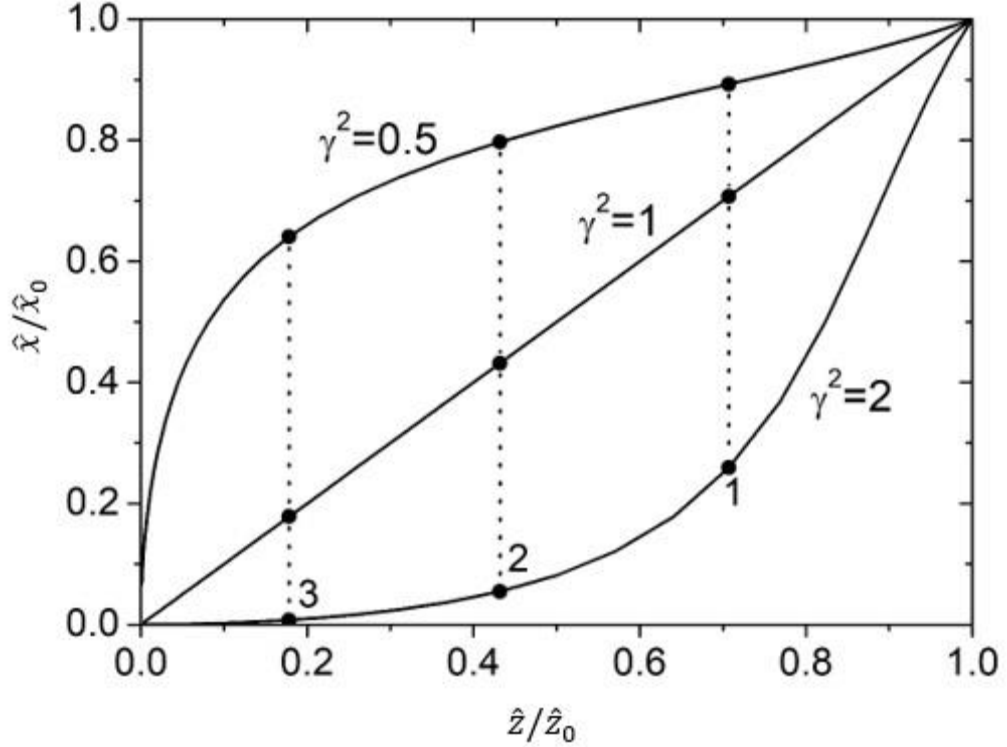

Fig. 3. Trajectories of particle motion for different values of $\gamma^2$. Points (1)-(3) on the curves correspond to the particle positions at various instants of time: (1) $\tau_z = 1.4$, (2) $\tau_z = 2.2$, (3) $\tau_z = 3.2$.

Not interested in the time dependence of the coordinates $\hat{z}$ and $\hat{x}$, it is easy to obtain from Eqs. (10) the formula for their relationship in the following form:

$$\ln\frac{\tan\hat{z}}{\tan\hat{z}_0} = \frac{1}{\gamma^2}\ln\frac{\tan\hat{x}}{\tan\hat{x}_0}. \tag{13}$$

Obviously, in this case, the relationship between the coordinates of the particle is determined by the values $\hat{z}_0$ and $\hat{x}_0$ and the ratio of intensities $\gamma^2$. As expected, the trajectories calculated from relation (13) under identical initial conditions $\hat{z}_0 = \hat{x}_0 = 0.9\pi/2$ coincide with the corresponding curves in Fig. 3.

To justify these analytical results, we have performed numerical simulations of Eqs. (8a)-(8b). The spatio-temporal dynamics of particles obtained is shown in Fig. 4. One can see that the particle trajectories (Fig. 3) given by the analytical solution of truncated Eqs. (10a)-(10b) qualitatively match the projection of the curves to the spatial plane in Fig.4. However, the simulations discover more complicated damped oscillatory dynamics of particles in the vicinity of the zero-force point.

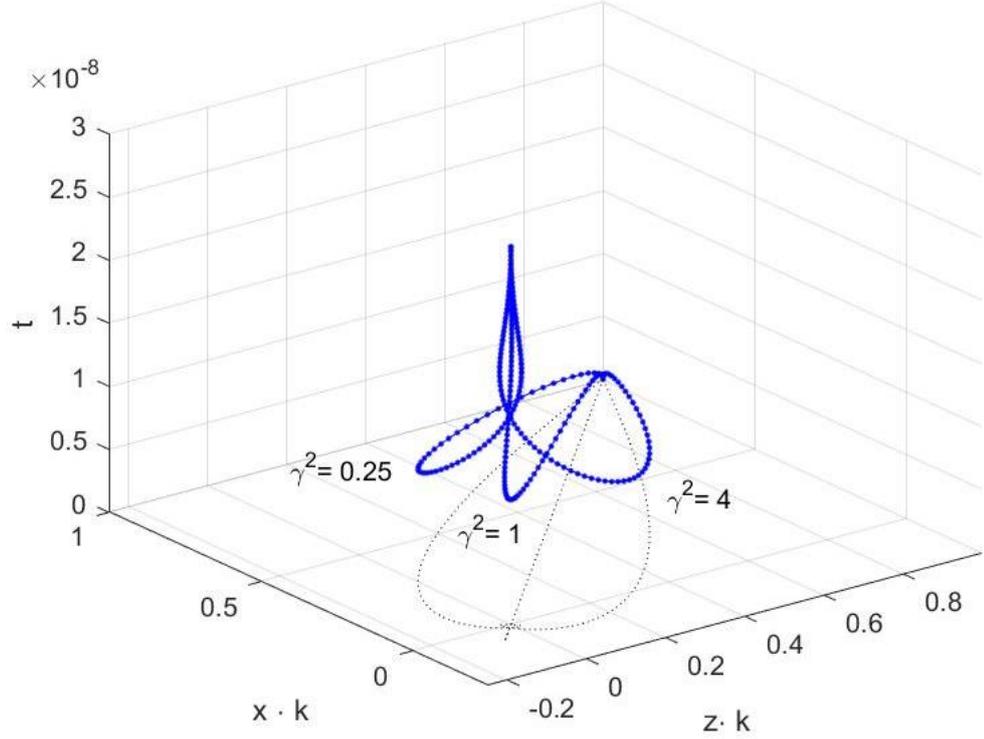

Fig. 4. Spatio-temporal dynamics of particles for different values of $\gamma^2$ obtained in numerical simulations using Eqs. (8a)-(8b) for the following parameter values: $kR = 1, \rho = 1, \eta = 10^{-2}, I_0 = 100, \Delta = 0, n = 1.33, \alpha = 0.12$.

### 4.2. The case of small particles $kR \ll 1$ at $\gamma^2 = 1$ and $\cos\Delta = \pm 1$

Given that for small particles, $u_0 \approx u_\Delta \approx 1$, from Eqs. (10) we can find the stationary equations for the connection between the particle coordinates:

$$\frac{d\hat{z}}{\sin\hat{z}} = \pm \frac{d\hat{x}}{\sin\hat{x}}, \tag{14}$$

where the signs "$\pm$" correspond to $\cos\Delta = \pm 1$.

Solutions of Eqs. (14) can be written as

$$\tan\frac{\hat{z}}{2} = \tan\frac{\hat{z}_0}{2} \begin{cases} \cot\dfrac{\hat{x}_0}{2} \cdot \tan\dfrac{\hat{x}}{2}, & \cos\Delta = 1, \quad (15a) \\ \tan\dfrac{\hat{x}_0}{2} \cdot \cot\dfrac{\hat{x}}{2}, & \cos\Delta = -1. \quad (15b) \end{cases}$$

As noted above, when the sign of the function $f(\Delta) = \cos\Delta$ changes, the interference pattern shifts by $\pi$. Therefore, when replacing $\hat{x}_0 \to \hat{x}_0 + \pi$ and $\hat{x} \to \hat{x} + \pi$, Eq. (15a) turns to Eq. (15b).

### 5. Conclusion

On the basis of the Newton equations, we have studied transportation of a spherical transparent particle under the influence of the gradient forces $\vec{F}_z$ and $\vec{F}_x$ in the interference field of two orthogonal standing waves of laser radiation $E_z \sim \cos 2kz$ and $E_x \sim \cos 2kx$. Such a problem statement, due to symmetry conditions, is the easiest to analyze the formation dynamics of a two-dimensional particle matrix similar to a molecular crystal, which was experimentally realized in [21]. It is shown that in the case of incoherent waves $E_z$ and $E_x$, the gradient forces depend only on a single ("their own") coordinate, $F_z(z)$ and $F_x(x)$, whereas in the case of coherent waves they are functions of both coordinates $F_z(z,x)$ and $F_x(z,x)$. The radiation intensity has two modulation components with the periods $\Lambda_0 = \pi/k$ and $\Lambda_\Delta = \pi/(k\sin(\pi/4))$. For the relatively large particle sizes $kR \gtrsim 1$, we have obtained the expressions for the particle polarizability, which are determined by the Bessel functions, $u_0 \sim J_{3/2}(2\pi R/\Lambda_0)$ and $u_\Delta \sim J_{3/2}(2\pi R/\Lambda_\Delta)$. The critical values of the particle radii $R_0$ are found, so that for $\cos\Delta = 0$, $\vec{F}_z = \vec{F}_x = 0$ regardless of the position of the particle on the interference radiation pattern $I_0 \sim (\cos(2kz), \cos(2kx))$. It is shown that the obtained values of $R_0$ differ from those following from the quantum theory of scattering [8] within less than 10%. In particular, for plastic particles in water under the action of radiation with $\lambda_0 = 1.064$ μm and for the first two zeros of the Bessel function $J_{3/2}(2\pi R/\Lambda_0)$, we have obtained $R_0 = 0.286$ μm and $0.492$ μm. In the coherent case when $\cos\Delta \neq 0$, the condition $\vec{F}_z \approx \vec{F}_x \approx 0$ is realized for relatively larger particle sizes, $kR \approx 5.45$.

It is shown that the initial system of truncated Newton equations in the case of incoherent waves $E_z$ and $E_x$ (for $\cos\Delta = 0$) splits into two independent equations that allow analytical solutions. For coherent waves ($\cos\Delta \neq 0$), we have obtained the analytical solutions for a number of special cases and the trajectories of the particle's motion to steady states.

The considered motion and localization of small particles by the gradient forces are the elementary processes of the formation of various spatial configurations from such particles, in particular concentration lattices, crystal-like structures, etc. Moreover, we suggest that the similar regularities can be observed and used in other physical settings, e.g., for manipulation of particles in dusty plasmas [26, 27].

**References**

[1] G.A. Askar'yan, Action of the gradient of an intense electromagnetic field beam on electrons and atoms, Sov. Phys. JETP 15 (1962) 1088-1090.

[2] V.S. Letokhov, Narrowing of the Doppler width in a standing wave, JETP Lett. 7 (1968) 272-275.

[3] A. Ashkin, Atomic-beam deflection by resonance-radiation pressure, Phys. Rev. Lett. 25 (1970) 1321-1324. https://doi.org/10.1103/PhysRevLett.25.1321

[4] A. Rohrbach, E.H.K. Stelzer, Optical trapping of dielectric particles in arbitrary fields, J. Opt. Soc. Am. A 18 (2001) 839-853. https://doi.org/10.1364/JOSAA.18.000839

[5] P. Zemánek, A. Jonáš, M. Liška, Simplified description of optical forces acting on a nanoparticle in the Gaussian standing wave, J. Opt. Soc. Am. A 19 (2002) 1025-1034. https://doi.org/10.1364/JOSAA.19.001025

[6] A.A. Afanas'ev, A.N. Rubinov, Yu.A. Kurochkin, S.Yu. Mikhnevich, I.E. Ermolaev, Localisation of spherical particles under the action of a gradient force in an interference field of laser radiation, Quant. Electron. 33 (2003) 250-254. https://doi.org/10.1070/QE2003v033n03ABEH002395

[7] A.A. Afanas'ev, A.N. Rubinov, S.Yu. Mikhnevich, I.E. Ermolaev, Four-wave mixing in a liquid suspension of transparent dielectric microspheres, J. Exp. Theor. Phys. 101 (2005) 389-400. https://doi.org/10.1134/1.2103207

[8] J. Lekner, Force on a scatterer in counter-propagating coherent beams, J. Opt. A: Pure Appl. Opt. 7 (2005) 238-248. https://doi.org/10.1088/1464-4258/7/5/005

[9] M.M. Burns, J.-M. Fournier, J.A. Golovchenko, Optical binding, Phys. Rev. Lett. 63 (1989) 1233-1236. https://doi.org/10.1103/PhysRevLett.63.1233

[10] M.M. Burns, J.-M. Fournier, J.A. Golovchenko, Optical matter: Crystallization and binding in intense optical fields, Science 249 (1990) 749-754. https://doi.org/10.1126/science.249.4970.749

[11] J. Arlt, V. Graces-Chavez, W. Sibbett, K. Dholakia, Optical micromanipulation using a Bessel light beam, Opt. Commun. 197 (2001) 239-245. https://doi.org/10.1016/S0030-4018(01)01479-1

[12] C. Das, P. Chaudhuri, A.K. Sood, H.R. Krishnamurthy, Laser-induced freezing in 2-d colloids, Current Science 80 (2001) 959-971.

[13] M.P. MacDonald, L. Paterson, W. Sibbett, K. Dholakia, P.E. Bryant, Trapping and manipulation of low-index particles in a two-dimensional interferometric optical trap, Opt. Lett. 26 (2001) 863-865. https://doi.org/10.1364/OL.26.000863

[14] A.A. Afanas'ev, V.M. Katarkevich, A.N. Rubinov, T.Sh. Efendiev, Modulation of particle concentrations in the interference laser radiation field, J. Appl. Spectr. 69 (2002) 782-787. https://doi.org/10.1023/A:1021525517007

[15] A.N. Rubinov, V.M. Katarkevich, A.A. Afanas'ev, T.Sh. Efendiev, Interaction of interference laser field with an ensemble of particles in liquid, Opt. Commun. 224 (2003) 97-106. https://doi.org/10.1016/S0030-4018(03)01728-0

[16] P.W. Smith, A. Ashkin, W.J. Tomlinson, Four-wave mixing in an artificial Kerr medium, Opt. Lett. 6 (1981) 284-286. https://doi.org/10.1364/OL.6.000284

[17] D. Rogovin, S.O. Sari, Phase conjugation in liquid suspensions of microspheres in the diffusive limit, Phys. Rev. A 31 (1985) 2375-2389. https://doi.org/10.1103/PhysRevA.31.2375

[18] A.A. Afanas'ev, A.N. Rubinov, S.Yu. Mikhnevich, I.E. Ermolaev, Theory of stimulated concentration scattering of light in a liquid suspension of transparent microspheres, Opt. Spectr. 102 (2007) 106-111. https://doi.org/10.1134/S0030400X07010195

[19] I.S. Burkhanov, S.V. Krivokhizha, L.L. Chaikov, Stimulated concentration (diffusion) light scattering on nanoparticles in a liquid suspension, Quant. Electron. 46 (2016) 548-554. https://doi.org/10.1070/QEL15700

[20] I.S. Burkhanov, S.V. Krivokhizha, L.L. Chaikov, Stokes and anti-stokes stimulated Mie scattering on nanoparticle suspensions of latex, Opt. Commun. 381 (2016) 360-364. https://doi.org/10.1016/j.optcom.2016.07.020

[21] C.D. Mellor, C.D. Bain, Array formation in evanescent waves, Chem. Phys. Chem. 7 (2006) 329-332. https://doi.org/10.1002/cphc.200500348

[22] Y. Harada, T. Asakura, Radiation forces on a dielectric sphere in the Rayleigh scattering regime, Opt. Commun. 124 (1996) 529-541. https://doi.org/10.1016/0030-4018(95)00753-9

[23] T. Li, Physical Principle of optical tweezers, in Fundamental Tests of Physics with Optically Trapped Microspheres, Springer, 2013, p. 9-20. https://doi.org/10.1007/978-1-4614-6031-2

[24] I.S. Gradshteyn, I.M. Ryzhik, Tables of Integrals, Series, and Products, Elsevier, 2007.

[25] A.A. Afanas'ev, D.V. Novitsky, Applicability of the approximate Langevin equation for describing the motion of nanospheres in the field of a standing light wave, Opt. Spectr. 125 (2018) 944-947. https://doi.org/10.1134/S0030400X18120019

[26] V.E. Fortov, A.G. Khrapak, S.A. Khrapak, V.I. Molotkov, O.F. Petrov, Dusty plasmas, Phys. Usp. 47 (2004) 447-492. https://doi.org/10.1070/PU2004v047n05ABEH001689

[27] V.E. Fortov, A.V. Ivlev, S.A. Khrapak, A.G. Khrapak, G.E. Morfill. Complex (dusty) plasmas: Current status, open issues, perspectives, Phys. Rep. 421 (2005) 1-103. https://doi.org/10.1016/j.physrep.2005.08.007